\documentclass[proof]{pasj00}
\draft

\begin{document}
\SetRunningHead{Kawate et al.}{Center-to-limb variation of radio emission}

\title{Center-to-Limb Variation of Radio Emissions from Thermal-Rich and Thermal-Poor Solar Flares}

\author{
  Tomoko \textsc{Kawate} \altaffilmark{1}
  Ayumi    \textsc{Asai}\altaffilmark{2}
  and
  Kiyoshi  \textsc{Ichimoto}\altaffilmark{1}}
\altaffiltext{1}{Kwasan and Hida Observatory, Kyoto University \\
  17, Kitakazan Ohmine-cho, Yamashina-ku, Kyoto, 607-8471, Japan}
\altaffiltext{2}{Unit of Synergetic Studies for Space, Kyoto University \\
  17, Kitakazan Ohmine-cho, Yamashina-ku, Kyoto, 607-8471, Japan}

\email{kawate@kusastro.kyoto-u.ac.jp}


%


\maketitle

\begin{abstract}
A statistical analysis of radio flare events was performed by using the event list of Nobeyama 
Radioheliograph in 1996-2009. 
We examined center-to-limb variations of 17GHz and 34GHz flux by dividing the flare events 
into different groups with respect to the `thermal plasma richness' (ratio of the peak flux of
soft X-ray to non-thermal radio emissions) and the duration of radio bursts.
It is found that peak flux of 17 and 34GHz tend to be higher toward the limb for thermal-rich
flares with short durations. We propose that the thermal-rich flares, which are supposed to be 
associated with an efficient precipitation of high energy particles into the chromosphere, have 
a pitch angle distribution of non-thermal electrons with a higher population along the flare loop.

\end{abstract}

\section{Introduction}
Non-thermal radio emissions observed in the impulsive phase of solar flares are produced by the gyrosynchrotron mechanism, which depends on a number of physical parameters such as 
electron energy spectra, their pitch angle distribution, 
magnetic field strength, angle between line of sight and  the magnetic field (viewing angle),  
and the number of electrons.
Therefore, it is difficult to determine those physical parameters uniquely only 
from the observed quantities of individual radio burst.
Statistical analyses of radio bursts by using a number of flare events 
provide us a way to find mutual relationships
between different quantities, and thus are useful to restrict the possible 
domain of those physical quantities of the radio source.
The pitch angle distribution of accelerated electrons is of a crucial importance 
for the problem of particle acceleration in solar flares.
A clue to know the pitch angle distribution of accelerated particles
could be obtained from the center-to-limb variations of observed radio 
emissions, since relativistic electrons trapped in flare loops
emit the radio waves to the direction of their velocity, and the viewing angle effect, 
i.e., center-to-limb variation of the flare emission, can be related to  the pitch angle distribution 
of accelerated electrons.

A number of authors have performed statistical studies of flare parameters and investigated their center-to-limb variations,
e.g., Dodson et al. (1954), Akabane(1956), Kundu (1959), Hervey (1964), Scalise (1970), Matsuura \& Nave (1970), 
Croom \& Powell (1971), Castelli \& Barron (1977), Kosugi (1985), Pogodin et al. (1996), and Silva \& Valente (2002).
Some authors classified radio bursts based on the properties of their light curves, i.e.,  single/multiple bursts,
gradual rise and fall type burst and so on. Dodson et al. (1954) showed that gradual rise and fall bursts
in 2.8GHz has a high concentration toward the central part of the solar disk. 
In frequencies lower than 10GHz where the gyrosynchrotron emission is optically thick, a weak center-to-limb variation of the peak intensity was observed, i.e., the peak intensity gets slightly larger toward the limb. Takakura \& Scalise (1970) and Ramaty (1969) argued that such center-to-limb
variation can be explained with a bipolar magnetic field configuration containing non-thermal 
electrons with anisotropic pitch angle distribution, while the Razin Suppression in ambient plasmas 
taken into account to suppress the degree of the center-to-limb variation.

In 19 and 17GHz where the plasma is mostly optically thin, Croom \& Powel (1971) and Kosugi (1985) found no significant center-to-limb 
variation of the peak flux. 
Contrary, Silva \& Valente (2002) found a significant center-to-limb
variation of gyrosynchrotron emission with a higher intensity toward the limb
in 3.1, 5.2, 8.4, 11.8, 19.6, 35.0 and 50GHz, and that the variation gets more
prominent in higher frequency ranges, i.e., in optically thin emission produced
by higher energy electrons, than in lower frequency ranges. 
Vestrand et al. (1987) reported that hard X-ray emission with energy higher than 25keV has a center-to-limb variation by which 
the intensity is larger and the power law index is harder near the limb than these near the disk center.  Datlowe et al. (1977), however, found that
the emission is isotropic and there are no significant center-to-limb variation in X-ray intensity  below 20keV.

Since the angle between line of sight and the magnetic field near the footpoints of flare loops gets larger toward the limb,
flare events near-the-limb 
are to be brighter than the disk center events, if the footpoints are the major sources of the radio emission.
In the theoretical view of Takakura \& Scalise (1970), a center-to-limb 
variation of radio intensity is actually expected with a magnetic dipole field trapping non-thermal electrons,
while its degree is dependent on the pitch angle distribution of the electrons.
The fact that there found no significant center-to-limb variation in the frequency range below 17GHz
was attributed  to
the complexity and variety of the magnetic field configuration among flares (Kosugi 1985).

The aim of this paper is
to find the insights on the pitch angle distribution of non-thermal electrons by investigating
 center-to-limb variations of radio intensity for different types of flares. 
To this end, we performed a statistical analysis by categorizing the flares in their duration,
i.e., the impulsiveness.
We use the radio data at two frequencies in optically thin ranges, 17 and 34GHz, to confirm the 
frequency dependence of the center-to-limb variation (Silva \& Valente 2002).
We also focus on the soft X-ray emission associated with the radio bursts.
During the impulsive phase of solar flares, non-thermal electrons
 precipitate into the chromosphere,
which leads to explosive plasma heating and pressure enhancement there.
As a result, hot thermal plasmas emitting soft X-rays rapidly fill flare
loops (these are so-called ``chromospheric evaporation").
Since non-thermal electrons should play a significant role in generation of
the thermal plasma as inferred from the empirical rule, so-called the
Neupert Effect (Neupert 1968), the soft X-ray emission could be an indicator
of the energy injection by non-thermal particles into the chromosphere.
Furthermore, since the precipitation rate of non-thermal electrons would be related with the pitch angle distribution, the 
thermal plasma richness, i.e., the soft X-ray flux relative to the non-thermal flux, will provide a measure of the 
pitch angle distribution of electrons.

We use the radio burst data obtained by the Nobeyama Radioheliograph (NoRH) (Nakajima et al. 1994), and 
 study the center-to-limb variation of intensity and other flare parameters. 
GOES data are also used to categorize the flares in their thermal richness. 
In Section 2, we describe the dataset of flare events used in this study. In Section 3, we define parameters for use of
the statistical analysis, and describe their statistical properties.
The `thermal emission index'  is introduced and relation to the other flare parameters are shown. In Section 4, we calculate correlation coefficients  
between each parameter and the heliocentric angle, and in Section 5 we discuss the results of the Section 4.

\section{Data Selection}
NoRH is a two dimensional interferometer, observing the intensity and 
circular polarization in 17GHz from 1992, and 34GHz intensity from 1995. 
With a data coverage of 97\% in Japanese daytime, about 2000 flare events are observed and
cataloged in the event list during the period from 1992 to 2009.

We use the event lists of the NoRH (http://solar.nro.nao.ac.jp/norh/html/event/ for strong events and 
http://solar.nro.nao.ac.jp/norh/html/eventw/ for weak events).
Our dataset for the statistical analysis covers the period from 1996 April 1 to 2009 December 31, during which both 17GHz and 34GHz data are available.
The event lists contain the duration of 
17GHz burst, peak flux of 17GHz and 34GHz, GOES peak
flux, and position on the solar disk, of each radio burst, while the
center-to-limb variation of the sensitivity of the instrument due to the finite 
beam size of each antenna is already corrected for the radio flux (Hanaoka et al. 1994).
We selected flare events whose peak flux at 17GHz is larger than 20 sfu, thus there is no selection
effect caused by the center-to-limb variation of sensitivity of the instrument. 
We discarded 31 events that have the maximum brightness temperature
larger than 10$^8$K.  Those flares is so bright compared with the quiet
disk with the brightness temperature of about 10$^4$K that the image
synthesis often fails due to the insufficient dynamic range of the
instrument.  For those events, neither the position of the flare on the
solar disk nor the correct peak flux can be determined, since these are
inferred from the quiet solar disk.

\begin{figure}
  \begin{center}
    \FigureFile(160mm,80mm){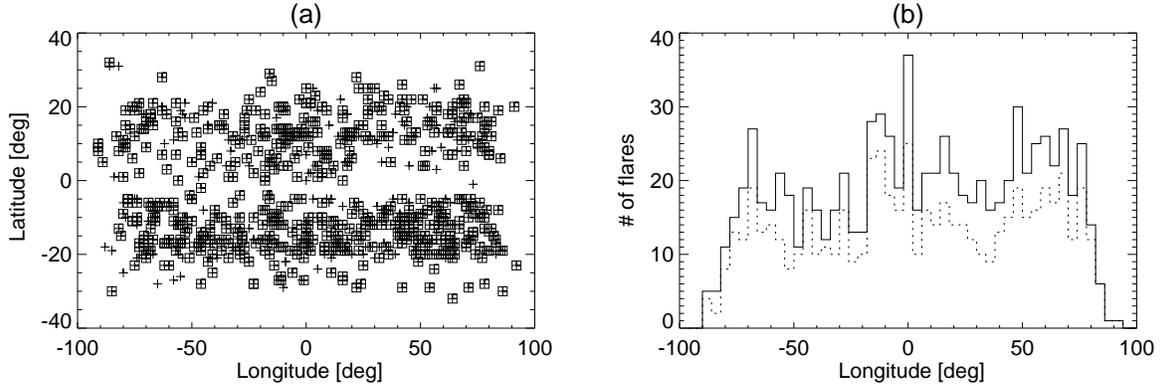}
     \caption{(a) Distribution of the flare events on the sun. All events are shown by $+$, while $\square$ shows events that have a data of the GOES peak flux. (b) Histogram of the events. All events are shown in solid line, and events with GOES data are shown in dotted line.}\label{fig:f1}
  \end{center}
\end{figure}

As a result, we selected 818 events, while 640 events among them have associated GOES flux data. 
Positions of the events on the solar disk are shown in Figure \ref{fig:f1} (a) with the symbol 
$+$, while events that have records of associated GOES peak flux are indicated by $\square$.
Figure \ref{fig:f1} (b) shows the 
histogram of events on the heliocentric longitude. A histogram for all radio events is shown by solid line, and that for events with the GOES data 
is shown by dotted line. The events of our dataset are widely distributed over the disk.

\section{Definition of Flare Parameters and their Statistics}
\subsection{Duration}
The duration of bursts are defined as the difference between the start time and the end time of the flare, which are determined from the 
the correlation value in 17GHz among different antennas (see, ``NOBEYAMA RADIOHELIOGRAPH CATALOG of EVENTS" (NRO 2002) for definition). Histogram of the duration is 
shown in logarithmic scale in Figure \ref{fig:f2} (a). The mean duration is 757 sec, and the standard deviation is 0.67 in logarithmic scale.
We categorize the events into three groups according to the duration; i.e., events with durations shorter than 400 sec (273 events), longer than 400 sec and shorter than 1800 sec (271 events), 
and longer than 1800 sec (274 events).
They are refereed as `{\it short}', `{\it medium}', and `{\it long}' events hereafter.

\subsection{Peak Flux of 17GHz and 34GHz}
Figure \ref{fig:f2} (c) and (d) show the histogram of the maximum flux of 17GHz and 34GHz, respectively. 
The black line is the histogram for 
all events, the gray line is the histogram for the events with a duration longer than 400 sec ({\it medium} and {\it long} events), 
and the dotted line 
is for the events longer than 1800 sec ({\it long} events). In Figure \ref{fig:f2} (c),  a power law fitting of the histogram in the 
interval of $1.5 < \log I_{\rm 17GHz} < 2.5$ is shown.
The power law index is $-$0.70, which leads to $dN/dI \propto {I_{\rm 17GHz}}^{-1.70}$ with $N$ the number of flares that have a peak intensity in unit bin of intensity $I_{\rm 17GHz}$. This result ($\gamma$=$-$1.70 where $\gamma$ is power law index) is in good
agreement with results by Kosugi (1985)($\gamma$=$-$1.7 in 17GHz), Akabane (1956)($\gamma$=$-$1.8 in 3GHz), Kundu (1959)($\gamma$=$-$1.5 in 2.8 and 9.4GHz), and Kakinuma et al. (1969) ($\gamma$=$-$1.8 in 1, 2, 3.75, 9.4GHz). 
The power law indices for all flares, {\it medium}+{\it long} events, and {\it long} events are $-$0.70, $-$0.36 and $-$0.22, respectively, i.e., 
as the duration of flare increases, the power law index of the histogram becomes larger (harder).
This means that {\it long} events  
have higher population in a large maximum flux in 17GHz. 
The same plot for the maximum flux of 34GHz is shown in Figure \ref{fig:f2} (d). 
The power law indices for the flares with all flares, {\it medium}+{\it long} events, and {\it long} events are $-$0.75, $-$0.57 and $-$0.48, respectively,
and they have the same tendency with 17GHz.

\subsection{Power Law Index of Radio Flux}
\begin{figure}
  \begin{center}
    \FigureFile(160mm,160mm){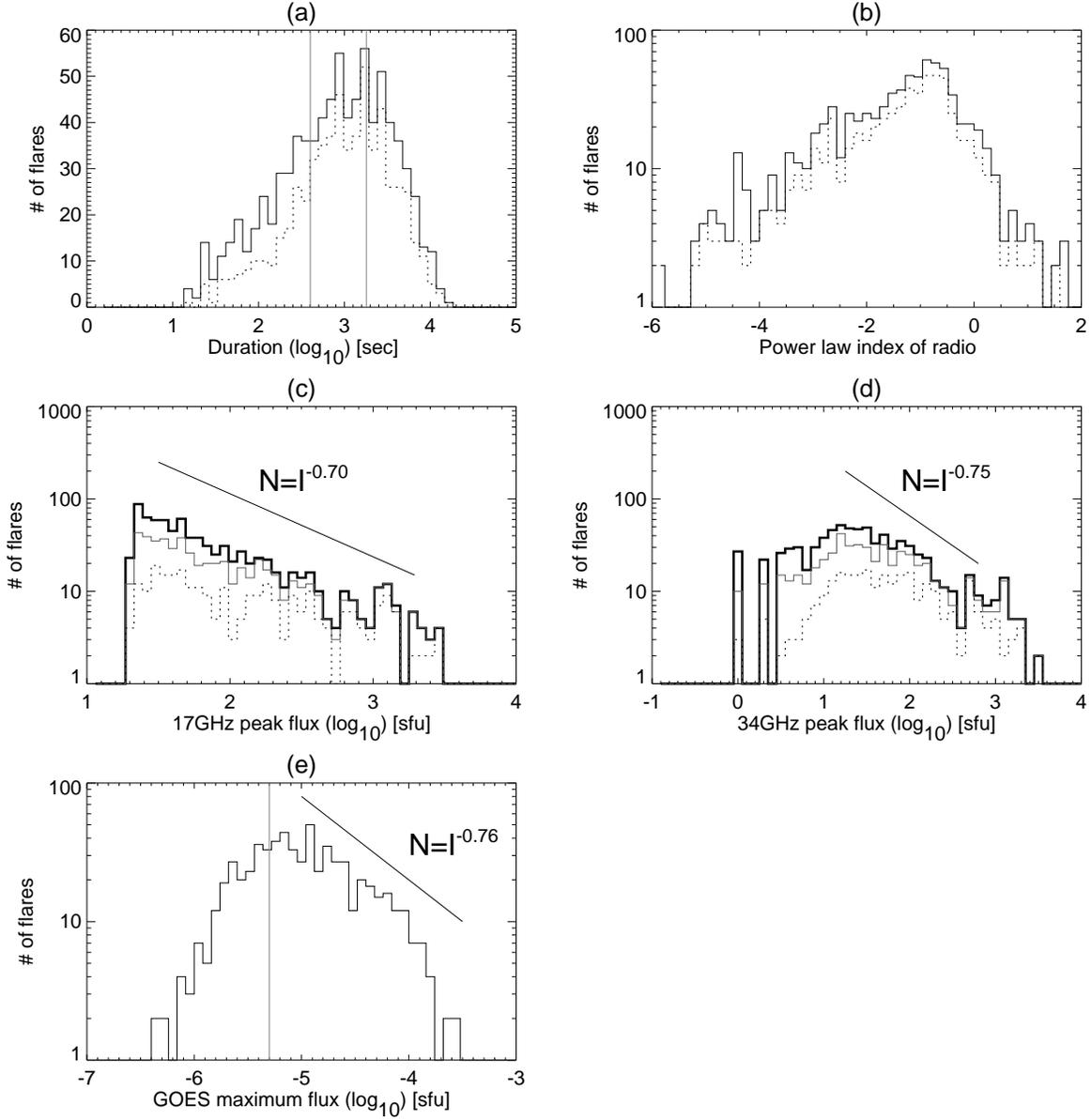}
     \caption{(a) Histogram of the duration of all events. All events are shown in solid line, and events with GOES data are shown 
in dotted line. The vertical thin line is on the 400 sec and 1800 sec. (b)Histogram of the power law index of radio bursts. (c)Histogram of the maximum 17GHz flux on each event and (d) 
34GHz flux on each event. The black thick line means $0<$duration$<400$s, the gray line means $400<$duration$<1800$s, whereas the dotted line means duration $>1800$s. (e)Histogram of the maximum GOES flux on each event. The vertical thin line shows the threshold of C5.0.}\label{fig:f2}
  \end{center}
\end{figure}

The power law index of each radio burst is defined as the ratio of peak flux at 34GHz and 17GHz. The Histogram of the power law 
index are shown in Figure \ref{fig:f2} (b). The mean value is $-$1.48, and standard deviation is 1.38. Although the peak fluxes of 17 
and 34GHz did not always take place at the same time,
94\% of the events have simultaneous peak in 17 and 34GHz.  

\subsection{GOES Flux}
The GOES flux cataloged in the NoRH event lists is the maximum soft X-ray flux of flares in 1\AA -8\AA. The histogram of GOES flux is shown in
Figure \ref{fig:f2}(e). As Figure \ref{fig:f2}(e) shows, small GOES events below C5.0 (vertical thin line in Figure \ref{fig:f2}(e)) 
are sometimes missed, since
the associated peak flux in 17GHz was smaller than the threshold of our event selection
By fitting to the histogram in interval between $10^{-5}$  and $10^{-4}$ W/m$^2$, 
We obtain a power law distribution of flare events as $dN/dI \propto {I_{\rm GOES}}^{-1.76}$.
This is in good agreement with the result by Hudson (1991), who showed $dN/dI \propto$ ${I_{\rm GOES}}^{-1.8}$.

\subsection{Thermal Emission Index}
\begin{figure}
  \begin{center}
    \FigureFile(160mm,160mm){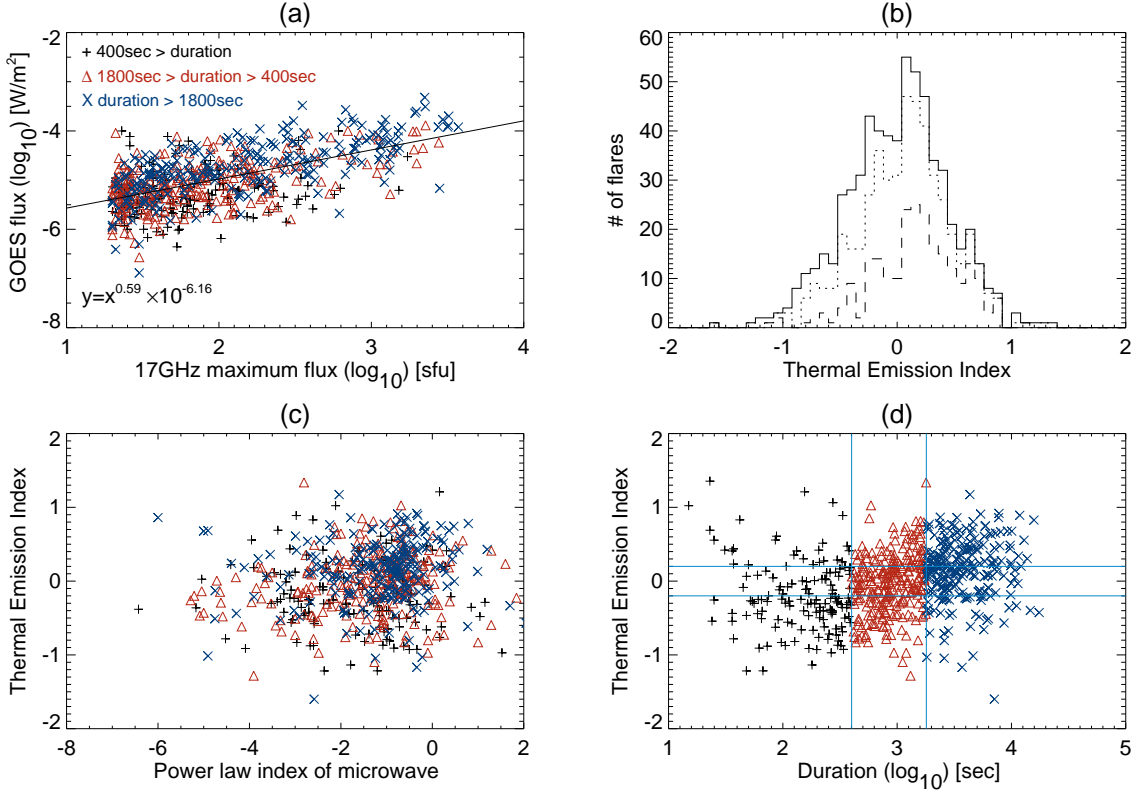}
     \caption{(a)Scattering plot.of maximum 17GHz flux and GOES flux. flares whose 
duration is shorter than 400 sec duration are marked with $+$, from 400 sec to than 1800 sec are marked with $\triangle$, and 
longer than 1800 sec are marked with $\times$. (b) Histogram of Thermal Emission Index. The solid line means the duration  $>0$, the dotted line means the duration $>400$, and the dashed line means the duration $>1800$. (c) Scattering plot of TEI and power spectral index. Symbols indicate the duration of the flares. (d) Scattering plot of 
TEI and duration of each event. Symbols indicate the duration of the flares.}\label{fig:f3}
  \end{center}
\end{figure}

A scatter plot of GOES flux and 17GHz peak flux is shown in Figure \ref{fig:f3} (a). 
There is a positive correlation between them with a  correlation coefficient of 
0.59, and the 95\% confidence interval of [0.54, 0.64]. Different symbols correspond to the different durations of flares; 
`$+$' , `$\triangle$', and `$\times$' show {\it short}, {\it medium}, and {\it long} events, respectively. In the following plots, we use the same symbols 
for the different durations.

A liner regression for the scatter plot is shown by a straight line in Figure \ref{fig:f3} (a). 
The flares above this line can be regarded as `thermal-rich', while those below which are  `thermal-poor'.
We define the ``Thermal Emission Index(TEI)" of flares as the vertical distance from this line, i.e., with the following formula, to evaluate the 
thermal plasma richness of each event;
\begin{eqnarray}
{\rm TEI} = \log _{10} \frac{I_{\rm GOES}}{{I_{\rm 17GHz}}^{0.59} } + 6.16
\end{eqnarray} 
An event with TEI=0 has a `typical' ratio of GOES peak flux and 17GHz peak flux. 
If we have two events with the same 17 GHz fluxes but TEIs are 0 and 1,
respectively, the latter has higher soft X-ray flux with the GOES class
of 1.
 We show the histogram of the TEI in Figure \ref{fig:f3} (b) and find that the standard deviation of TEI is 0.45.
The colors of lines in Figure \ref{fig:f3} (b) show different duration of flares as in Figure \ref{fig:f2} (c).

There is no correlation between TEI and 17GHz peak flux 
by the definition of TEI. Also no significant correlation between TEI and 
34GHz is found (0.05 [$-$0.01, 0.12]). We show the relation between TEI and power law index of radio emissions
in Figure \ref{fig:f3} (c). The correlation coefficient is 0.11 [0.04, 0.17], so that they have a very weak positive 
correlation in a sense that thermal-rich flares have slightly harder power law index. The relation between TEI and duration is shown in Figure \ref{fig:f3} (d),
in which we find a significant positive correlation (0.22 [0.15,0.28]) among them.
Why {\it long} events tend to be thermal-rich may be attribute to 
 that {\it long} events usually consist of 
multiple radio bursts, while the GOES flux is emitted from the thermal plasma accumulated by 
those energy depositions. Thus, even with a similar peak flux of 17GHz, GOES X-ray can have 
a larger peak flux for the longer events.

\section{Center-to-Limb Variations}
The center-to-limb variation of the radio emission is the main subject of this section.
Here, flare events are divided into three groups with respect to TEI, i.e. TEI$<-0.2$ (thermal poor), $-0.2<$TEI$<0.2$, and $0.2<$TEI (thermal rich),
and also into three groups with respect to the duration as mentioned in previous section ({\it short}, {\it medium}, and {\it long} events). 
Division lines into the 9 groups in total are shown in the scatter plot for TEI versus duration in Figure \ref{fig:f3} (d), while
the numbers of events belonging to each group are given in Table \ref{tab:evnum}.
\begin{table}[htbp]
\begin{center}
\begin{tabular}{|c|c|c|c|c|}
\hline
TEI$\backslash$duration [sec]    & $<$ 400   & 400 - 1800  & 1800 $<$  & all       \\ \hline \hline
$<$ $-$0.2     & {\bf 83}        & 77          & 42        & 202       \\ \hline
$-$0.2 - 0.2   & 34        & {\bf 105}         & 80        & 219       \\ \hline
0.2 $<$      & 27        & 71          & {\bf 111}       & 209       \\ \hline
all          & 144       & 253         & 233       & 630       \\
\hline
\end{tabular}
  \caption{Number of events in each group.}
  \label{tab:evnum}
\end{center}
\end{table}

\begin{figure}
  \begin{center}
    \FigureFile(160mm,160mm){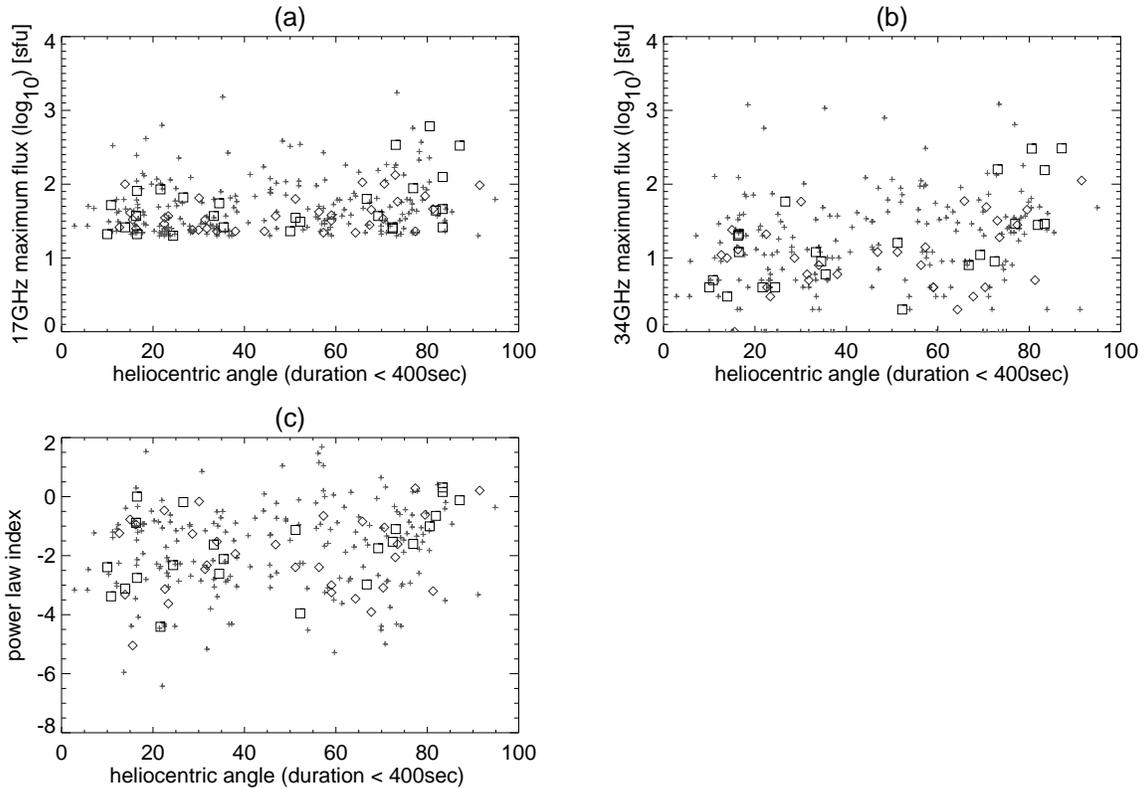}
     \caption{Center-to-limb variation of (a) 17GHz peak flux, (b) 34GHz peak flux, (c) and power law index for {\it short} events.
 The symbol $+$ shows flares with TEI $<$ $-$0.2 (thermal-poor), $\diamond$ for $-$0.2 $<$ TEI $<$ 0.2, and $\square$ for $<$ 0.2 (thermal-rich).}\label{fig:f5}
  \end{center}
\end{figure}

The correlation coefficients of the 17GHz peak flux and the heliocentric angle for each group 
are shown in Table \ref{tab:17-hcd}, and those of
the 34GHz peak flux and heliocentric angle in Table \ref{tab:34-hcd}. In these tables, it is remarkable that 
the thermal-rich events (TEI $>0.2$) with {\it short} and {\it medium} events have a significant positive
 correlation between the peak flux and the heliocentric angle in both 17GHz and 34GHz.
We also show the correlation between the power law index of radio flux and the heliocentric angle in Table \ref{tab:pli-hcd}.
It is found again that the power law index shows a significant positive correlation with the heliocentric
angle for the thermal-rich and {\it short}-{\it medium} events, so that the relative intensity of 34GHz against 17GHz gets higher toward the limb.
Correlation plots for 17GHz peak flux, 34GHz peak flux, and the power law index versus the heliocentric angle for {\it short} events 
are shown in Figure \ref{fig:f5}, (a), (b), and (c), respectively. 
The symbol `$+$' is a flare with TEI $<$ $-$0.2 (thermal-poor), `$\diamond$' for $-$0.2 $<$ TEI $<$ 0.2, and `$\square$' for $<$ 0.2 (thermal-rich). 

\begin{center}
\begin{table}[htbp]
\begin{center}
\begin{tabular}{|c|c|c|c|c|}
\hline
TEI$\backslash$duration [sec]  & $<$ 400             & 400 - 1800          & 1800 $<$            & all         \\ \hline \hline
$<$ $-$0.2   & 0.13 [$-$0.01, 0.26]  & $-$0.17 [$-$0.34, 0.01] & $-$0.06 [$-$0.27, 0.16] &$-$0.02 [$-$0.12, 0.07]       \\ \hline
$-$0.2 - 0.2 & {\bf 0.42} [0.10, 0.67]   & 0.01 [$-$0.19, 0.19]  & 0.01 [$-$0.13, 0.30]  &  0.07 [$-$0.06,0.20]       \\ \hline
0.2 $<$    & {\bf 0.45} [0.08, 0.71]   & 0.08 [$-$0.31, 0.16]  &  0.08 [$-$0.11, 0.26] & 0.04 [$-$0.10,0.17]       \\ \hline
all        & 0.19[$-$0.07,0.30]    & $-$0.08 [$-$0.19,0.03]  & 0.08 [$-$0.04,0.19]   & 0.04 [$-$0.12,0.07]       \\
\hline
\end{tabular}
\end{center} 
  \caption{Correlation coefficient between maximum flux of 17GHz and heliocentric angle.}
  \label{tab:17-hcd}
\end{table}
\end{center}

\begin{center}
\begin{table}[htbp]
\begin{center}
\begin{tabular}{|c|c|c|c|c|}
\hline
TEI$\backslash$duration [sec]   & $<$ 400             & 400 - 1800          & 1800 $<$            & all         \\ \hline \hline
$<$ $-$0.2    & 0.14 [0.01, 0.27]   & $-$0.09 [$-$0.26, 0.01] & $-$0.01 [$-$0.22, 0.21] & 0.02 [$-$0.07, 0.12]       \\ \hline
$-$0.2 - 0.2  & {\bf 0.35} [0.02, 0.61]   & 0.01 [$-$0.18, 0.20]  &  0.14 [$-$0.08, 0.35] & 0.10 [$-$0.03,0.23]       \\ \hline
0.2 $<$     & {\bf 0.63} [0.33, 0.81]   & $-$0.13 [$-$0.35, 0.11]   & $-$0.02 [$-$0.21, 0.16] & 0.00 [$-$0.13,0.14]       \\ \hline
all         & 0.21[0.09,0.32]     & $-$0.05 [$-$0.16,0.07]   & 0.06 [$-$0.06,0.19]   & 0.06 [$-$0.01,0.012]       \\
\hline
\end{tabular} 
\end{center}
  \caption{Correlation coefficient between maximum flux of 34GHz and heliocentric angle.}
  \label{tab:34-hcd}
\end{table}
\end{center}
\begin{center}
\begin{table}[htbp]
\begin{center}
\begin{tabular}{|c|c|c|c|c|}
\hline
TEI$\backslash$duration [sec]  & $<$ 400             & 400 - 1800          & 1800 $<$             & all         \\ \hline \hline
$<$ $-$0.2   & 0.14 [0.00, 0.27]   & 0.04 [$-$0.14, 0.22]  & 0.11 [$-$0.11, 0.32]   & 0.03 [$-$0.00, 0.18]       \\ \hline
$-$0.2 - 0.2 & 0.15 [$-$0.20, 0.46]  & $-$0.05 [$-$0.24, 0.14] &  0.15 [$-$0.07 0.36]   & 0.05 [$-$0.08,0.18]       \\ \hline
0.2 $<$    & {\bf 0.45} [0.09, 0.71]   & $-$0.17 [$-$0.39, 0.07] & $-$0.21 [$-$0.38, $-$0.02] & $-$0.08 [$-$0.21,0.06]       \\ \hline
all        & 0.17[0.05,0.29]     & 0.00 [$-$0.12,0.11]  & 0.00 [$-$0.12,0.11]   & 0.05 [$-$0.01,0.12]      \\
\hline
\end{tabular} 
\end{center}
  \caption{correlation coefficient between power law index of radio and heliocentric angle.}
  \label{tab:pli-hcd}
\end{table}
\end{center}
 
\begin{figure}
  \begin{center}
    \FigureFile(80mm,80mm){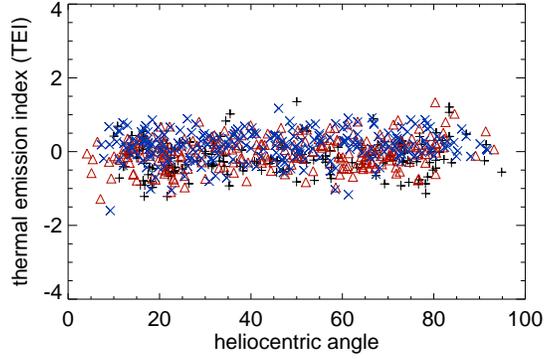}
     \caption{Center-to-limb variation of Thermal Emission Index. Symbols indicate the duration of the flares; $+$ is flares whose duration is shorter than 400 sec duration, $\triangle$ longer than 400 sec and shorter than 1800 sec, and $\times$ longer than 1800 sec.} \label{fig:f4}
  \end{center}
\end{figure}

As for the TEI versus the heliocentric angle, the correlation coefficient is 0.10 [0.04, 0.17] for all events,
0.10 [$-$0.02, 0.21] for {\it short}, 0.15 [0.05, 0.27] for {\it medium} and 0.10 [$-$0.02,0.22] for {\it long} events (Figure \ref{fig:f4}). 
The weak positive correlation found in the {\it medium} events is originated in the center-to-limb variation of 
GOES flux in our dataset. 
We suppose that this tendency is produced in the selection of events by means of the NoRH 
signal, but the exact cause is not understood.

\section{Discussion}

As shown, in Table \ref{tab:17-hcd}, there is no center-to-limb variation in 17GHz peak flux when we take
all events as a whole. 
This result agrees with Kosugi (1985). However, the peak flux of non-thermal radio bursts for flares with large TEI and short duration 
has a significant center-to-limb variation in both 17 and 34GHz ,i.e., are higher for near-the-limb events. 

It is found in Table \ref{tab:17-hcd}, \ref{tab:34-hcd}, and Figure 
\ref{fig:f5} (a)(b) that the higher frequency has stronger center-to-limb variation, as also can be 
 seen in the power law index that gets harder toward the limb.
This result agrees with the result of Silva \& Valente (2002), i.e., the higher the frequency, the 
higher the tendency of the center-to-limb variation of flux.

The significant center-to-limb variation of radio flux found in thermal-rich events could be interpreted as 
follows; since thermal plasma is produced by the precipitation of non-thermal
electrons to the chromosphere,
it is supposed that high TEI flares have a pitch angle distribution
with high population in small angle (along the magnetic field)
so that an efficient precipitation of electrons to the chromosphere takes place.
Under such pitch angle distribution, large number of electrons
including lower energy ones
can travel to the mirroring points located in a lower height in the corona,
i.e. footpoints of the flare loop, where magnetic field is strong.
Thus, lower energy electrons efficiently emit the gyrosynchrotron
and the footpoints get the dominant sources of the flare emission.
As the consequence, the observed radio emission gets brighter toward the limb.

The reason why only {\it short} events have the center-to-limb variation of radio flux seems to be that, for {\it long} events, 
soft X-ray flux is created by multiple non-thermal bursts, so that there is no one to one correspondence between 
the radio peak flux and the soft X-ray peak, thus TEI has no significant meaning and probably does not represent the 
pitch angle distribution of the non-thermal electrons.
We also think that, in {\it long} events, electrons tend to be trapped 
in coronal loop with a pitch angle distribution to be more perpendicular to 
the magnetic field, so that emission comes from looptop and there is less center-to-limb variation.

In conclusion, our statistical study of flare bursts presents the first observational evidence on that the efficiency 
of generating the thermal plasma is related to the pitch angle distribution of the non-thermal electrons of 
solar flares.

\bigskip
\section*{Acknowledgment}
The authors thank Drs. K.Shibasaki, H. Nakajima, M. Shimojo of National Astronomical  Observatory of Japan, Dr. S. Masuda of Nagoya University,
Dr. T. Yokoyama of University of Tokyo, and Dr. T. Minoshima in JAMSTEC for valuable discussions.
This work is supported by the Grant-in-Aid for the Global COE Program ``The Next Generation of Physics, Spun from
Universality and Emergence" from the Ministry of Education, Culture, Sports, Science and Technology of Japan.
and partially carried out by the joint research program of the
Solar-Terrestrial Environment Laboratory, Nagoya University.


\end{document}